\documentclass[twocolumn]{aastex631}

\usepackage{mathtools}
\begin{document}

\title{On the presence of metallofullerenes in fullerene-rich circumstellar envelopes}

\correspondingauthor{R. Barzaga}
\email{rbarzaga@iac.es}

\author[0000-0002-9827-2762]{R. Barzaga}
\affiliation{Instituto de Astrof\'{\i}sica de Canarias, C/ Via L\'actea s/n, E-38205 La Laguna, Spain}
\affiliation{Departamento de Astrof\'{\i}sica, Universidad de La Laguna (ULL), E-38206 La Laguna, Spain}
\author[0000-0002-1693-2721]{D. A. Garc\'{\i}a-Hern\'andez}
\affiliation{Instituto de Astrof\'{\i}sica de Canarias, C/ Via L\'actea s/n, E-38205 La Laguna, Spain}
\affiliation{Departamento de Astrof\'{\i}sica, Universidad de La Laguna (ULL), E-38206 La Laguna, Spain}

\author[0000-0001-6253-6343]{S. D\'{\i}az-Tendero}
\affiliation{Departmento de Qu\'{\i}mica, M\'{o}dulo 13, Universidad Aut\'{o}noma de Madrid, 28049 Madrid, Spain}
\affiliation{Institute for Advanced Research in Chemical Science (IAdChem), Universidad Aut\'{o}noma de Madrid, 28049 Madrid, Spain}
\affiliation{Condensed Matter Physics Center (IFIMAC), Universidad Aut\'{o}noma de Madrid, 28049 Madrid, Spain}

\author[0000-0003-3529-0178]{SeyedAbdolreza Sadjadi}
\affiliation{Research Center for Theoretical and Experimental Physics, Chemistry and Space Sciences, Genius Development and ScienceTech Future Co
Ltd, Hong Kong (SAR), PR China}
\affiliation{Laboratory for Space Research, Faculty of Science, Department of Physics, The University of Hong Kong, Hong Kong (SAR), PR China}

\author[0000-0002-3011-686X]{A. Manchado}
\affiliation{Instituto de Astrof\'{\i}sica de Canarias, C/ Via L\'actea s/n, E-38205 La Laguna, Spain}
\affiliation{Departamento de Astrof\'{\i}sica, Universidad de La Laguna (ULL), E-38206 La Laguna, Spain}
\affiliation{Consejo Superior de Investigaciones Cient\'{\i}ficas (CSIC), Spain}

\author[0000-0002-3753-5215]{M. Alcami}
\affiliation{Departmento de Qu\'{\i}mica, M\'{o}dulo 13, Universidad Aut\'{o}noma de Madrid, 28049 Madrid, Spain}
\affiliation{Institute for Advanced Research in Chemical Science (IAdChem), Universidad Aut\'{o}noma de Madrid, 28049 Madrid, Spain}
\affiliation{Instituto Madrile\~no de Estudios Avanzados en Nanociencia (IMDEA-Nano), Campus de Cantoblanco, Madrid 28049, Spain}

%
%
%
%
%



\begin{abstract}

The presence of neutral C$_{60}$ fullerenes in circumstellar environments has
been firmly established by astronomical observations as well as laboratory
experiments and quantum-chemistry calculations. However, the large variations
observed in the C$_{60}$ 17.4$\mu$m/18.9$\mu$m band ratios indicate that either
additional emitters should contribute to the astronomical IR spectra or there
exist unknown physical processes besides thermal and UV excitation.
Fullerene-based molecules such as metallofullerenes and fullerene-adducts are
natural candidate species as potential additional emitters, but no specific
species has been identified to date. Here we report a model based on quantum-chemistry
calculations and IR spectra simulation of neutral and charged endo(exo)hedral metallofullerenes, showing
that they have a significant contribution to the four strongest IR bands
commonly attributed to neutral C$_{60}$. These simulations may explain the large range of
17.4$\mu$m/18.9$\mu$m band ratios observed in very different fullerene-rich
circumstellar environments like those around planetary nebulae and
chemically peculiar R Coronae Borealis stars. Our proposed model also reveals that
the 17.4$\mu$m/18.9$\mu$m band ratio in the metallofullerenes simulated IR spectra mainly depends on the metal abundances,
ionization level, and endo/exo concentration in the circumstellar envelopes. We
conclude that metallofullerenes are potential emitters contributing to the
observed IR spectra in fullerene-rich circumstellar envelopes. Our simulated IR
spectra indicate also that the James Webb Space Telescope has the potential to
confirm or refute the presence of metallofullerenes (or even other
fullerene-based species) in circumstellar enviroments.

\end{abstract}



\section{Introduction} \label{sec:intro}
The detection of the most common fullerene species (C$_{60}$) in Planetary Nebulae (PNe) and diverse astrophysical environments have raised the exciting possibility that other more complex fullerene-based molecules (e.g., metallofullerenes, multishell fullerenes, fullerene-adducts) might be ubiquitous in space and continue to be serious candidates to explain several astrophysical phenomena like the unidentified infrared bands (UIRs), the UV-bump, and the diffuse interstellar bands (DIBs), among others \citep[see][for a review]{Kwok2016}. This idea was reinforced in 2015 when the fullerene cation (C$_{60}^+$) was established as the only DIB carrier known to date \citep{Campbell2015}.

The presence of neutral C$_{60}$ in space is deduced from the detection of its four strongest mid-IR emission bands (those at $\sim$7.0, 8.5, 17.4, and 18.9 $\mu$m)\footnote{The mid-IR features of the fullerene cation (C$_{60}^+$) have only been detected in the reflection nebula NGC 7023 \citep{Berne2013}.}. The neutral C$_{60}$ mid-IR bands have been (mainly) detected in the circumstellar environments of young PNe \citep[e.g.,][]{Cami2010,Garcia2010} but also in other types of circumstellar environments like those around R Coronae Borealis (RCB) stars \citep{Garcia2011a}.

Recent experimental studies combined with quantum-chemical calculations demonstrate also that fullerenes would react with metal atoms and molecules (e.g., polycyclic aromatic hydrocarbons; PAHs), forming a rich family of fullerene-based molecules such as endo(exo)hedral metallofullerenes and fullerene-PAH adducts \citep[e.g.,][]{Dunk2013}. These fullerene derivatives may still be excited by UV photons, emitting through the same IR vibrational modes as empty C$_{60}$ cages. For example, laboratory work shows that the strongest mid-IR features of fullerene-PAH adducts are strikingly coincident with those from neutral C$_{60}$, suggesting that fullerene-based molecules may contribute to the four C$_{60}$ mid-IR features observed in fullerene-containing circumstellar environments \citep{Garcia2013}.

In fact, \cite{Brieva2016} recently made a detailed comparison of their laboratory inferred C$_{60}$ emission band strengths with the astrophysical data on fullerene-rich circumstellar environments, showing that the observed strengths cannot be explained in terms of fluorescent or thermal emission alone. They concluded that the C$_{60}$ emission ratios of 17.4$\mu$m/18.9$\mu$m observed imply that they have a contribution of other emitters, or that physical processes other than thermal or UV excitation affect the C$_{60}$ vibrational modes\footnote{\cite{Brieva2016} also demonstrate that, in contrast to other C$_{60}$ band ratios (e.g., those involving the 7.0 and 8.5 $\mu$m features), the 17.4$\mu$m/18.9$\mu$m band ratio should remain almost constant independently of the C$_{60}$ excitation model assumed (fluorescence or thermal models).}. Here we present a theoretical study based on quantum-chemistry
calculations of endo(exo)hedral metallofullerenes (both neutral and charged), showing that they significantly contribute to the four IR bands generally attributed to neutral C$_{60}$ and may explain the large range of the C$_{60}$ 17.4$\mu$m/18.9$\mu$m band ratio observed in very different circumstellar environments like those around PNe and RCB stars. Our simulated IR spectra indicate that the James Webb Space Telescope (JWST) has the potential to confirm or refute the presence of metallofullerenes in circumstellar enviroments.

\section{Computational Details} \label{sec:comput}
Quantum chemical calculations on the molecular geometry and spectroscopic properties of twenty-eight neutral and charged metallofullerenes have been performed in the framework of Density Functional Theory (DFT). In particular geometry optimization was carried out at the B3LYP/6-31G(d) level \citep{Becke1983,Lee1988,Ditchfield1971} with the Gaussian 16 code \citep{g16,gv6}. Accuracy benchmarking of this level of theory has been previously reported\citep{Robledo2014,Wang2017}. Spin multiplicity ($2S+1$) has been considered in the calculations to a maximum of $S=5/2$, according to the metal multiple charged states. The chemical formulas of these species are [M-C$_{60}$]$^{0/+}$ (exohedral) and [M@C$_{60}$]$^{0/+}$ (endohedral) where M = Li, K, Na, Ca, Mg, Fe, Ti\footnote{We consider the most abundant metals and/or well known to be present in the interstellar medium.}. All the geometries used has been characterized as local minimum, with non-imaginary frequencies observed. In order to simulate the vibrational spectra of these species the harmonic oscillator approximation was assumed and, subsequently, harmonic frequencies were adjusted applying a double-scaling-factors scheme to account for anharmonicity, vibro-rotational couplings, etc. \citep{Trujillo2022}. This approach guaranteed an error $<$ 2\% in the calculated frequencies of C$_{60}$, used as benchmarking system, with respect to the experiments \citep{Kern2013}. The total mixture spectra of the metallofullerenes were modelled by the simple addition of each IR intensity. In all the spectra, the peak intensity has been simulated by a Gaussian function of FWHM = 5 cm$^{-1}$, which is similar to the astronomical observations. The throughout procedure here implemented to simulate the intensity, accounting for the metal atoms concentration/abundance, is presented in the Appendix \ref{sec:spectra}. The atomic charges were obtained from electron density partitioning described by the Quantum Theory of Atoms in Molecules (QTAIM), employing the AimAll software \citep{Bader1990,AimAll}. Total relative energies have been computed with respect to the most stable structure, thus, a zero value represented it.

\section{Results and Discussion} \label{sec:results}
The simulated IR ($\sim$5-50 $\mu$m) spectra in the next two subsections have been constructed from the total mixture spectrum procedure (Sect. \ref{sec:comput}). In order to identify the distinctive spectral features of metallofullerenes, Figs. \ref{fig:f1} and \ref{fig:f2} include the four strongest features of C$_{60}$ \citep{Kern2013,Sadjadi2022}, according to our own calculations (Sect. \ref{sec:comput}). The blue dashed lines represent the simulated C$_{60}$ features with a height describing the intensity of the corresponding peak. Note that the intensity of the C$_{60}$ features was scaled equally to the amount of metallofullerenes; i.e., assuming that the same amount of C$_{60}$ molecules and metallofullerenes exist. The total weighted spectrum approach, however, is applied in the third subsection; for further details see Appendix \ref{sec:spectra}. Further information about the simulated IR spectra for all the metallofullerene species (28) will be publicly available for the astronomical community in a forthcoming paper (Barzaga et al.).

\subsection{Exofullerenes}
\begin{figure*}[ht!]
\centering
\includegraphics[scale=0.9]{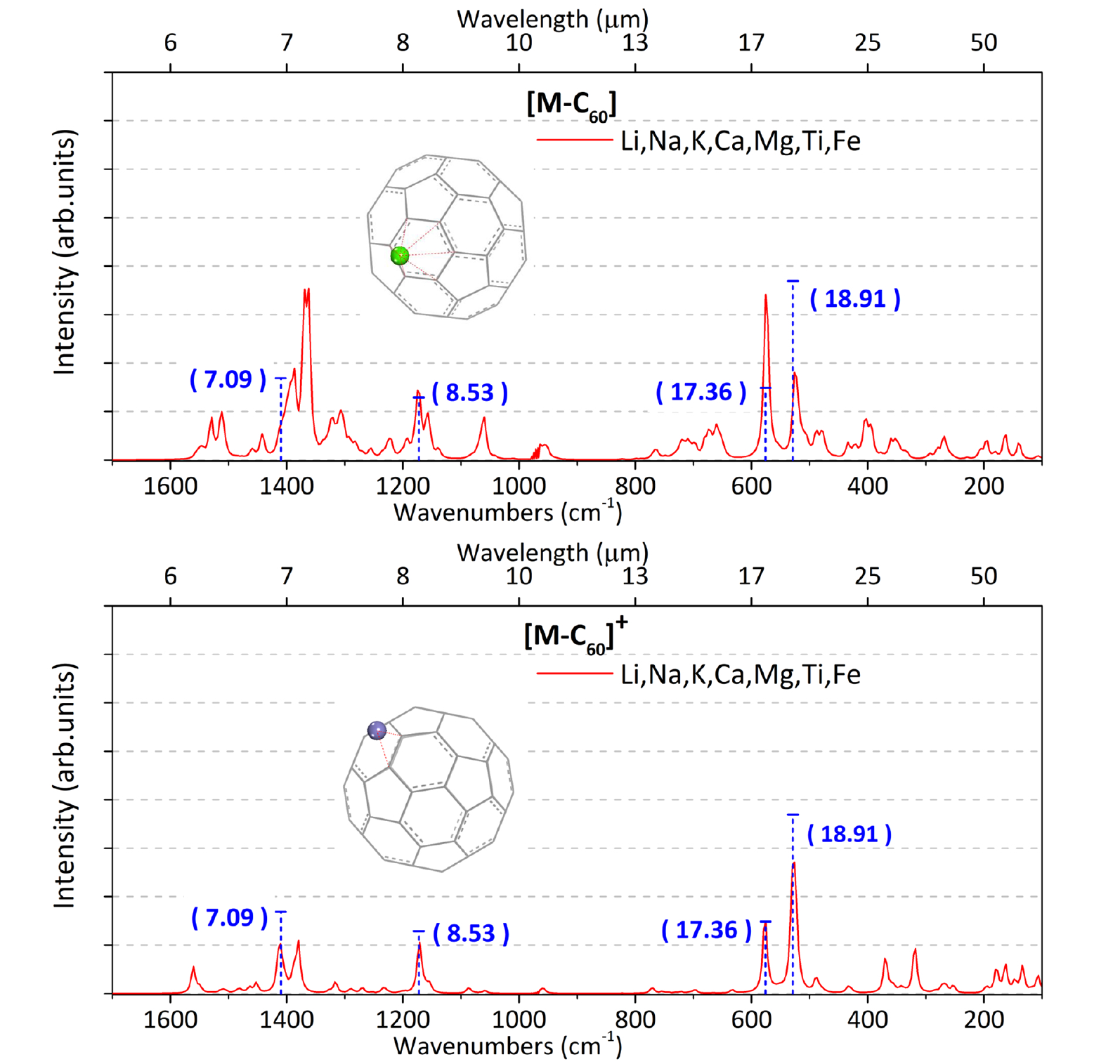}
\caption{DFT simulated IR ($\sim$5-50 $\mu$m) spectra of the mixture of metalloexofulerenes. Top: Seven neutral [M-C$_{60}$], Bottom: Seven charged [M-C$_{60}$]$^{+}$. Blued dashed lines represent the features of seven pristine C$_{60}$ molecules, whose intensity is described by the height of the lines. The insets display examples of the most common structures obtained.
\label{fig:f1}}
\end{figure*}
Figure \ref{fig:f1} shows the mixture of exofullerenes spectra, neutral and charged separately; C$_{60}$ features (in blue) are also included for comparison. The positions of the strongest features in both neutral and charged exofullerenes are coincident with those of C$_{60}$\footnote{The only exception is the $\sim$7.0 $\mu$m feature in the neutral exofullerenes, which is not among the strongest IR features. This feature is certainly present, but blended with the other emission components within the 6-9 $\mu$m spectral region.}. The neutral metalloexofullerenes [M-C$_{60}$] exhibit an IR spectrum richer than their charged counterparts; with a broadening and splitting of the IR features, mainly characterized by a modification of their intensity in comparison to C$_{60}$. Multiple features appear in the 6-9 $\mu$m region; specially, the highest intensity peak at 7.34 $\mu$m, which overcome the 7.09 $\mu$m C$_{60}$ feature. Similarly, above 14 $\mu$m, the spectral fingerprint of [M-C$_{60}$] is the emergence of several IR peaks accompanied with an inversion of the 17.4$\mu$m/18.9$\mu$m band ratio. The diversity of new 6-9 $\mu$m IR features in the [M-C$_{60}$] spectrum reflects the binding of the metal to the C$_{60}$ carbon cage, while the change in the 17.4$\mu$m/18.9$\mu$m ratio (which are the vibrational modes related to the cage stretching) it is likely due to a charge redistribution provoked also by the metal. The C$_{60}$ 17.4$\mu$m/18.9$\mu$m ratio displays values between 0.41 and 0.79, according to experiments and DFT calculations, respectively \citep{Kern2013,Sadjadi2022}; in our simulations such ratio for C$_{60}$ is 0.40 (blue dashed lines in Figure \ref{fig:f1}). However, the 17.4$\mu$m/18.9$\mu$m ratio in the [M-C$_{60}$] mixture spectrum is 1.89 (Fig. \ref{fig:f1}). Interestingly, the 17.4$\mu$m/18.9$\mu$m ratio is not inverted (0.55) in the charged metalloexofullerene [M-C$_{60}$]$^{+}$ mixture spectrum (Fig. \ref{fig:f1}), which is more characteristic of an electrostatic interaction \citep{Parker2010,Jaeger2004,Szczepanski2006} and non-covalent metal--C$_{60}$ forces \citep{Robledo2014}; i.e., the lack of new IR features in the 6-9 $\mu$m C-C stretching region together with a global spectral resemblance to pristine C$_{60}$.

\subsection{Endofullerenes}
\begin{figure*}[ht!]
\centering
\includegraphics[scale=0.9]{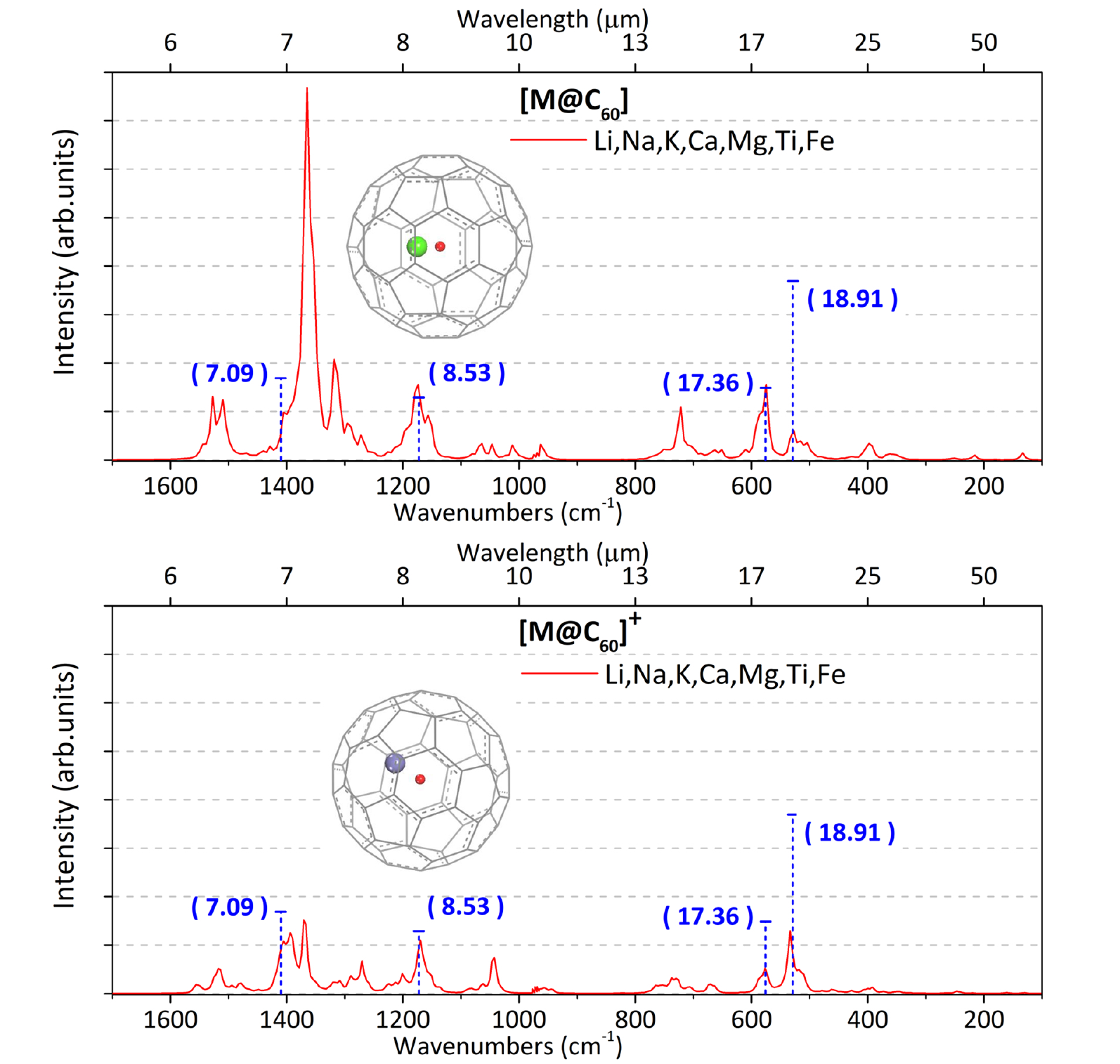}
\caption{DFT simulated IR ($\sim$5-50 $\mu$m) spectra of the mixture of metalloendofulerenes. Top: Seven neutral [M@C$_{60}$], Bottom: Seven charged [M@C$_{60}$]$^{+}$. Blued dashed lines retain the same description as in Figure \ref{fig:f1}. The insets display examples of the most common structures obtained but in this case a red dot depicts the centroid of the C$_{60}$ carbon cage.
\label{fig:f2}}
\end{figure*}
The neutral metalloendofullerenes [M@C$_{60}$] show their most intense feature at 7.32 $\mu$m because of the off-centered inclusion of the metal in the C$_{60}$ cage (Fig.\ref{fig:f2}). Such effect creates a strong asymmetry in the C$_{60}$ because carbon atoms trend to disrupt the cage in order to accommodate the metal. The perturbation of C-C stretching modes by the binding of the metal implies the whole cage instead of a section, like in the case of neutral metalloexofullerenes. This implies an important increment in the intensity of the IR features within the 6-9 $\mu$m region in comparison to pure C$_{60}$ (Fig.\ref{fig:f2}); even surpassing the IR features intensity for the neutral metalloexofullerenes mentioned above (Fig.\ref{fig:f1}). The strong broadening and splitting of the [M@C$_{60}$] mixture spectrum in the 6-9 $\mu$m region is again an indication of the metal-carbon cage binding\footnote{Similarly to the neutral exofullerenes, the $\sim$7.0 $\mu$m feature (being not among the strongest features) is present but blended with other higher intensity features inside the 6-9 $\mu$m spectral region.}. However, above 14 $\mu$m the spectral landscape of [M@C$_{60}$] considerably diminishes; specially the 18.9$\mu$m feature intensity, which results in a 17.4$\mu$m/18.9$\mu$m ratio of 2.55. In the case of charged metalloendofullerenes [M@C$_{60}$]$^{+}$ the reduction of the IR features intensity takes place over all the $\sim$5-50$\mu$m spectral region; particularly, in the 6-9$\mu$m C-C stretching region. Such intensity reduction likely implies a lack of charge transfer between the metal and C$_{60}$ carbon cage. In spite of this, the 17.4$\mu$m/18.9$\mu$m ratio (0.40) is almost identical to the one for C$_{60}$.

\begin{figure*}[ht!]
\centering
\includegraphics[scale=0.9]{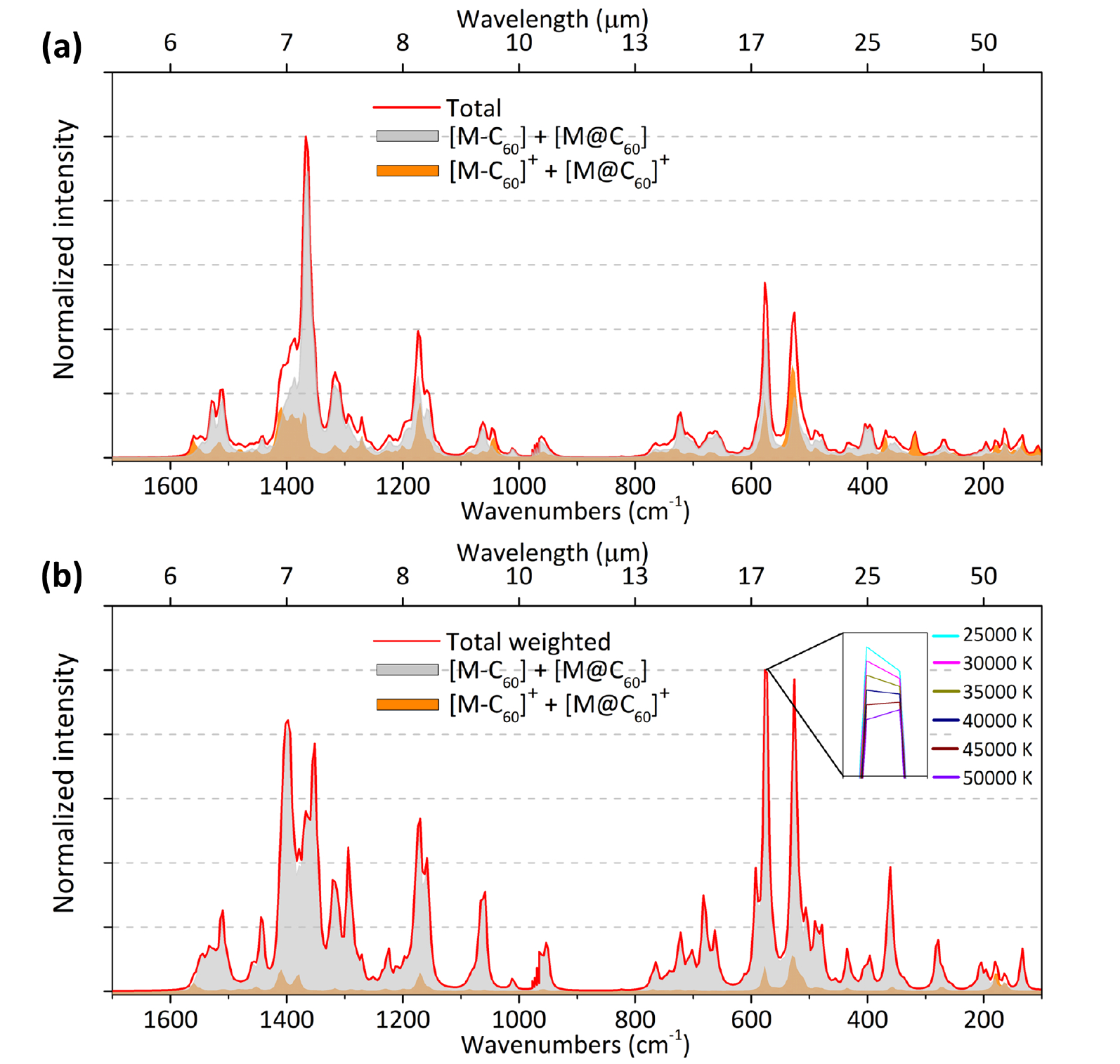}
\caption{Contributions of neutral and charged metallofullerenes to: (a) Total mixture spectrum from the twenty-eight metallofullerenes. (b) Total weighted spectrum from abundances and ion fraction. Neutral and charged species are denoted by a light-gray and orange filled areas, respectively. An inset highlights the T$_{eff}$ dependent change in the 17.4 $\mu$m feature intensity for the total weighted spectrum.
\label{fig:f3}}
\end{figure*}
\subsection{Charge vs Neutrality}
Both charged and neutral metallofullerenes would coexist in circumstellar envelopes since one of the reactants in the metallofullerenes formation, the metals, can be already ionized by the central star \citep[e.g.][]{Garcia2012}. In any circumstellar environment, however, the metals have a particular abundance, implying that reliable IR spectra simulations should mix both neutral and charged species but including the specific metal abundances. For this, we have constructed the total weighted mixture spectra in very different fullerene-rich circumstellar envelopes like those around PNe and RCB stars (see details in Appendix \ref{sec:spectra}). Figure \ref{fig:f3}a and b show, respectively, total and total weighted mixture spectra for the C$_{60}$-PNe case; the intensity has been normalized to ease the comparison. We note that the formation probability of endo- versus exo-fullerenes has not been considered because this would require the study of complex kinetic processes, which are out of the scope of this paper. We thus simply assume that both exo- and endo-fullerenes have the same fomation probability.

The total mixture spectra has the neutral metallofullerenes as the dominant contributor with an integrated spectral area of 68\% (see Fig.\ref{fig:f3}a). Surprisingly, this drastically changes for the total weighted mixture spectra (see Fig.\ref{fig:f3}b), where the neutral metallofullerenes completely dominate the spectrum, with an integrated spectral area of 94\%. The metal abundances and ionization fractions expected in C$_{60}$-PNe circumstellar environments are the two main factors producing such strong modifications in the total weighted mixture spectra. This is because in C$_{60}$-PNe: i) the metal abundance follows the order, Mg $>$ Fe $>>$ Na $>$ Ca $>$ K$\simeq$Ti $>$ Li
\citep[][see Appendix \ref{sec:spectra}]{Karakas2010,Karakas2018}; and ii) the most abundant metals (Fe, Mg) exhibit the lowest fractions of ionized atoms, around 5-10\%; even for the highest effective temperature $T_{eff}$ = 50,000 K observed in C$_{60}$-PNe \citep{Garcia2012,Otsuka2014}. Indeed, direct thermal effects over the IR intensity can be almost neglected; only very small deviations (a factor of 1.07; see the inset in Fig.\ref{fig:f3}b) are predicted for the narrow $T_{eff}$ range ($\sim$25,000-50,000 K) of C$_{60}$-PNe. The noticeable modification of the total weighted mixture spectra prove the importance of the metal abundances and ionization fractions in order to obtain reliable IR simulated spectra. The
spectral landscape in Figure \ref{fig:f3}b is altered in such a way that the
17.4$\mu$m/18.9$\mu$m ratio varies from 1.21 to 1.03 (total vs total weighted spectra). Thus, our results indicate that the 17.4$\mu$m/18.9$\mu$m ratio can be used to track down the presence of charged against neutral metallofullerenes.

Chemical stability is another key factor for the likely preponderance of neutral metallofullerenes. Table \ref{tab:tab1} displays the relative total energies ($E_{tot}^{rel}$, see also Sect.\ref{sec:comput}) of all metallofullerenes studied here. Energetically neutral metallofullerenes are the more stable species by an order of $\sim$7 eV, which can be attributed to the interaction between the metal and C$_{60}$, where the charge is compensated. Oppositely, charged metallofullerenes create an uncompensated interaction with the C$_{60}$ cage. Specially for [M@C$_{60}$]$^{+}$ since the positive charge of the metal is embedded on the carbon cage, which requires more energy to stabilize this charge surrounded by an almost neutral environment. Furthermore, in some cases the C$_{60}$ cage is slightly positively charged hindering even more the energetical stabilization.

\begin{deluxetable*}{cccccccccccccccc}
\tabletypesize{\small}
\tablewidth{2pt}
\tablecaption{Total relative energies in eV ($E_{tot}^{rel}$) and charge on metal ($q^{M}$) and C$_{60}$ cage ($q^{C_{60}}$). \label{tab:tab1}}
\tablehead{ & \multicolumn{3}{c}{[M@C$_{60}$]} && \multicolumn{3}{c}{[M-C$_{60}$]} && \multicolumn{3}{c}{[M@C$_{60}$]$^{+}$} && \multicolumn{3}{c}{[M-C$_{60}$]$^{+}$} \\
\cline{2-4} \cline{6-8} \cline{10-12} \cline{14-16}
\colhead{} & \colhead{$E_{tot}^{rel}$}  & \colhead{$q^{M}$} & \colhead{$q^{C_{60}}$} & \colhead{} & \colhead{$E_{tot}^{rel}$}  & \colhead{$q^{M}$} & \colhead{$q^{C_{60}}$} & \colhead{} & \colhead{$E_{tot}^{rel}$}  & \colhead{$q^{M}$} & \colhead{$q^{C_{60}}$} & \colhead{} & \colhead{$E_{tot}^{rel}$}  & \colhead{$q^{M}$} & \colhead{$q^{C_{60}}$}
}
\startdata
{  } Ca & 0 & +1.78 & -1.78 && 0.22 & +0.91 & -0.91 && 5.54 & +1.78 & -0.78 && 5.23 & +0.97 & +0.03 \\
{  } Mg & 0.34 & +0.36 & -0.36 && 0 & +0.80 & -0.80 && 6.06 & +1.09 & -0.09 && 5.34 & +0.90 & +0.10 \\
{  } Li & 0 & +0.91 & -0.91 && 0.07 & +0.91 & -0.91 && 7.03 & +0.91 & +0.09 && 5.12 & +0.93 & +0.07 \\
{  } Na & 0 & +0.93 & -0.93 && 0.21 & +0.90 & -0.90 && 5.37 & +0.89 & +0.11 && 5.04 & +0.93 & +0.07 \\
{  } K & 0 & +0.93 & -0.93 && 0.56 & +0.91 & -0.91 && 5.36 & +0.93 & +0.07 && 5.16 & +0.94 & +0.06 \\
{  } Fe & 1.41 & +1.07 & -1.07 && 0 & +0.41 & -0.41 && 6.89 & +1.04 & -0.04 && 5.61 & +0.93 & +0.07 \\
{  } Ti & 0 & +1.36 & -1.36 && 0.46 & +0.92 & -0.92 && 5.88 & +1.35 & -0.35 && 5.89 & +1.21 & -0.21 \\
\enddata

\end{deluxetable*}

On the other hand, once neutral metallofullerenes are formed, charged species can be generated via the photoionization reactions:
\begin{equation*}
  \mathrm{[M@C_{60}]}\xrightarrow{\mathrm{h}\nu}\mathrm{[M@C_{60}]^{+}}
\end{equation*}
\begin{equation*}
  \mathrm{[M-C_{60}]}\xrightarrow{\mathrm{h}\nu}\mathrm{[M-C_{60}]^{+}}
\end{equation*}

\begin{figure}[ht!]
\includegraphics[scale=0.9]{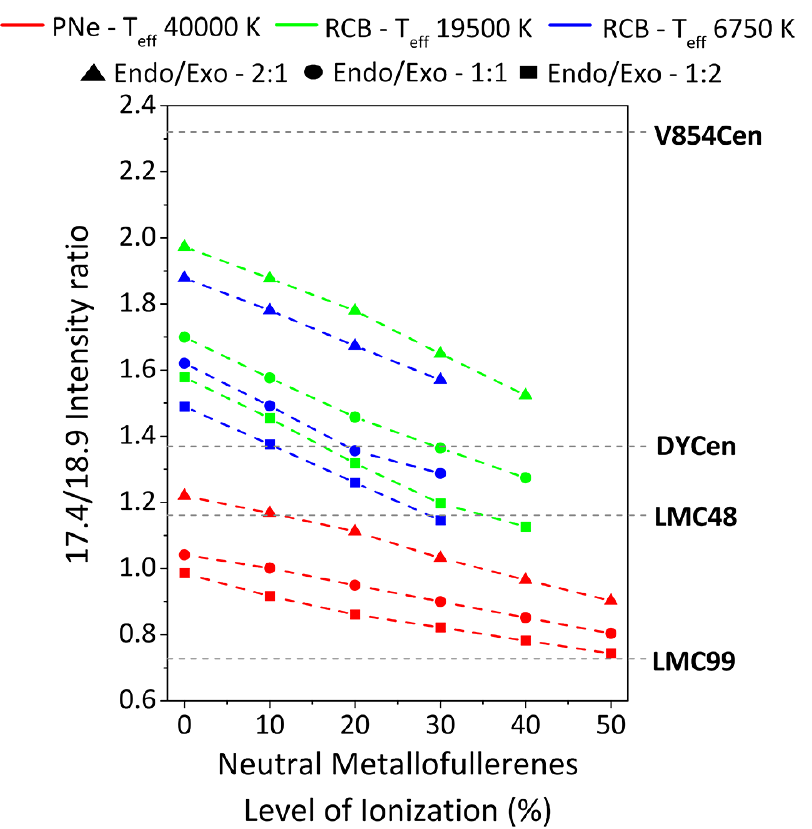}
\caption{The 17.4$\mu$m/18.9$\mu$m ratio vs neutral metallofulerenes level of ionization. The several colors and symbols correspond to different circumstellar envelopes (PNe and RCB) and exo/endo fullerenes concentrations, respectively (see legend). The horizontal dashed lines
indicate the 17.4$\mu$m/18.9$\mu$m ratios observed in two C$_{60}$-PNe \citep[LMC 48, LMC 99;][]{Garcia2012}, C$_{60}$-RCBs V854 Cen, DY Cen \citep[this work;][]{Garcia2011a}.
\label{fig:f4}}
\end{figure}

Both reactions need to accomplish two requirements: i) photons exciting the neutral metallofullerenes should have energies equal to the corresponding ionization potential; and ii) the production of charged metallofullerenes depends also on the amount of ionizing photons arriving. The first condition, the metallofullerenes ionization potential, can be inferred from the equation $IP =E^{total}_{rel}charged - E^{total}_{rel}neutral$ using the values presented in Table \ref{tab:tab1} between analogous species (e.g., [Ca@C$_{60}$] vs. [Ca@C$_{60}$]$^{+}$). This gives a range from $\sim$5 to 7 eV, which is an energy easily accessible for the photons emitted
by the C$_{60}$-PNe central stars (CS) \citep[O- and B-type stars;][]{Sternberg2003}. The second condition depends on the flux of ionizing photons leaving the central star, but this flux decays when the central star is cooler \citep{Sternberg2003}. We thus estimate that for a $T_{eff}=40,000$ K PN central star, a 50\% of the ionizing photons would produce charged metallofullerenes, while for the cooler ($<$20,000 K) RCB stars the flux decays gradually to 10\%. Such estimations allow us to observe the effect of charged metallofullerenes overproduction in the 17.4$\mu$m/18.9$\mu$m band ratio. Nevertheless, these are rough estimations since for the coolest RCB star V854 Cen ($T_{eff}=6,750$ K) almost no ionizing photons are produced according to our black body models. Figure \ref{fig:f4} displays the 17.4$\mu$m/18.9$\mu$m ratio as a function of the ionization level of neutral metallofullerenes. The data points associated to the curves of RCB stars were calculated according to Appendix \ref{sec:spectra}, using the adequate abundances for these stars.

The negative slope observed for the three types of circumstellar envelopes indicate that the 17.4$\mu$m/18.9$\mu$m ratio decreases when more charged metallofullerenes are
present; i.e., higher ionization level (Fig.\ref{fig:f4}). The difference between the
curves is a direct consequence of the metal abundances, which modify the initial amount of
neutral metallofullerenes. Thus, the PNe curves cover a 17.4$\mu$m/18.9$\mu$m range of
0.73-1.23, while the RCBs ranges are 1.14-1.88 (V854 Cen) and 1.12-1.97 (DY Cen). Thermal
effects mainly imply an increment of the ionization level although we actually cannot
discard possible temperature effects on the kinetics of metallofullerenes formation.
Strictly, temperature effects could be important in the transformation of endo- to
exo-fullerenes (or viceversa) but such study will be the subject of a future work. A
simple approach to infer the endo/exo concentration effect has been obtained from the
total amount of neutral or charged species. Therefore, either exo- or endo-fullerenes
concentration has been doubled for both neutral or charged species. The Figure
\ref{fig:f4} shows that an increment in the amount of endofullerenes (endo/exo 2:1) also
produces an increment in the 17.4$\mu$m/18.9$\mu$m ratio with respect to the endo/exo 1:1
concentration, while an exofullerenes increment (endo/exo 1:2) reduces such intensity
ratio. So, for the same ionization level, different 17.4$\mu$m/18.9$\mu$m ratios are
obtained depending on the endo/exo concentration. This behavior stressed the marked loss
of intensity in the 18.9$\mu$m feature provoked by metalloendofullerenes. However, the
general trend remains to be dominated by the presence of more neutral species against the
charged ones. As the level of ionization rises we would expect that the curves
corresponding to each circumstellar envelope tend to approximate each other.

\section{Astrophysical implications and concluding remarks}

As it was mentioned above, \cite{Brieva2016} have highlighted that the use of C$_{60}$
relative intrinsic strengths from laboratory data does not explain the large range
of 17.4$\mu$m/18.9$\mu$m band ratios ($\sim$0.2-1.2) observed in the fullerene-rich
circumstellar envelopes around PNe. One possible explanation is the presence of
additional emitters contributing to the observed IR spectra. Fullerene-based molecules
such as endo(exo)hedral metallofullerenes and fullerene-PAH adducts, among others, are
natural candidate species although, to our best knowledge, no additional emitter has been
identified to date.

Our DFT calculations and corresponding simulated IR spectra for metallofullerene species reveal that the 17.4$\mu$m/18.9$\mu$m band ratio mainly depends on the metal abundances, ionization level, and endo/exo concentration in the circumstellar envelopes. Interestingly, the theoretically predicted 17.4$\mu$m/18.9$\mu$m band ratios ($\sim$0.7-1.2) for C$_{60}$-PNe agree reasonably well with those observed in Galactic and extra-galactic C$_{60}$-PNe ($\sim$0.2-1.2) \citep{Garcia2011b,Brieva2016}; the match is quite exceptional for the Large Magellanic Cloud (LMC) PNe ($\sim$0.7-1.1) \citep{Garcia2011b}, which are the C$_{60}$-PNe with the metal abundances closest to the composition assumed in the construction of the total weighted mixture metallofullerene spectra. The latter is exemplified in Figure \ref{fig:f4} by the LMC PNe LMC 48 and LMC 99.

Furthermore, the consistency of our quantum-chemistry calculations goes beyond the
C$_{60}$-PNe and also explains the astonishing anomalous 17.4$\mu$m/18.9$\mu$m ratios
observed in the chemically peculiar RCB stars. The two known C$_{60}$-RCBs display a
18.9$\mu$m feature much weaker than the 17.4$\mu$m one \citep{Garcia2011a};
17.4$\mu$m/18.9$\mu$m ratios of $\sim$1.4 and 2.3 are observed in DY Cen and V854 Cen,
respectively (Fig.\ref{fig:f4}). The theoretically predicted 17.4$\mu$m/18.9$\mu$m band
ratios ($\sim$1.1-2.0) for the peculiar RCB compositions agree very well with the DY Cen
observation, while V854 Cen displays a 17.4$\mu$m/18.9$\mu$m ratio larger than the
predictions. This, however, is in agreement with the spectral analysis of their {\it
Spitzer} spectra, which shows that the 17.4$\mu$m feature is dominated by C$_{60}$
emission in DY Cen, while such feature in V854 Cen is a combination of C$_{60}$ and PAH
emission \citep{Garcia2011a}. The unusually large 17.4$\mu$m/18.9$\mu$m ratios observed in RCBs are naturally explained by a much larger contribution of neutral metallofullerenes, as expected from their lower T$_{eff}$ and weaker UV radiation fields.

In short, we conclude that metallofullerenes are potential emitters contributing to the
observed IR spectra in fullerene-rich circumstellar envelopes, providing for the first
time an explanation for the fundamental problem of the large range of
17.4$\mu$m/18.9$\mu$m band ratios observed.

We emphasize that a perfect spectral match of the simulated metallofullerene IR
spectra with the PNe and RCB {\it Spitzer} spectroscopic observations is not possible.
Apart from the four strongest neutral C$_{60}$ features, there is a no complete one to
one correspondence and we cannot discard the presence of others carriers (even other
fullerene-based species, see below) affecting the same IR features. However, from our
quantum-chemistry calculations it seems clear that the neutral metallofullerenes species
significantly contribute to the 6-9 $\mu$m C-C stretching region because of the
metal-carbon cage binding effect (see Sect.\ref{sec:results}). A similar spectral effect
would be thus expected for other fullerene-based neutral species. Indeed, a genuine
characteristic of fullerene-rich PNe environments with unusually high
17.4$\mu$m/18.9$\mu$m ratios like the LMC PNe analyzed here (see Fig.\ref{fig:f4}) is the
general presence of a broad and complex (with multiple components/peaks) IR feature at
6-9 $\mu$m \citep{Garcia2012}; something that strongly suggest the presence of fullerene-based neutral species such as neutral metallofullerenes, among others.

Finally, we note that most fullerene-rich circumstellar envelopes have been previously
observed at low resolution (R$\sim$120) by {\it Spitzer}, strongly limiting the detection
of the weaker (and more specific) mid-IR features of endo(exo)hedral metallofullerenes
(both neutral and charged; see Figs. \ref{fig:f1} to \ref{fig:f3}). Surprisingly, the only two C$_{60}$-MCPNe
observed at higher resolution (R$\sim$600; $\sim$10-20 $\mu$m) with {\it Spitzer} display
the presence of new IR emission features not previously observed in astrophysical
environments \citep{Garcia2011b}. Such new IR emission features are potentially due to
specific fullerene-based species such as metallofullerenes; most of them, however,
remained as tentative because of the low-quality (S/N$\sim$10 at the continuum) {\it
Spitzer} spectra \citep{Garcia2011b}. The JWST, with a much higher sensivity (and
spectral resolution) than {\it Spitzer}, has the potential to unambiguously detect the
spectral signatures of new IR emission features potentially due to metallofullerenes (or
even other fullerene-based species). Thus, high S/N ($>$100) and high-resolution
(R$\sim$2,400 on average) JWST spectroscopic observations of fullerene-rich envelopes
(e.g., C$_{60}$ PNe and RCBs) are strongly encouraged in order to confirm or refute the
presence of metallofullerenes in circumstellar environments.


\section{Acknowledgments}
 
We acknowledge support from the ACIISI, Gobierno de Canarias, and the European Regional Development Fund (ERDF) under a grant with reference PROID2020010051 as well as the State Research Agency (AEI) of the Spanish Ministry of Science and Innovation (MICINN) under grants PID2020-115758GB-I00 and PID2019-110091GB-I00. This article is based upon work from COST Action NanoSpace, CA21126, supported by COST (European Cooperation in Science and Technology).

\appendix
\section{Weighted Simulated Spectra} \label{sec:spectra}
The IR intensity $(I)$ can be defined as $I\propto(\partial \mu/\partial x)^{2}$ where $\partial \mu$ stands for the change in the dipole moment and $\partial x$ the displacement produced by the corresponding vibrational mode. Another factor that determines the peak intensity in the IR is the concentration ($n_{i}$) of $i$ molecules or species:
\begin{equation}\label{eqn:1}
I\propto\sum^{j}_{i}\left(\frac{\partial \mu}{\partial x}\right)^{2}n_{i}
\end{equation}
The summation indicates the contribution from species $i^{th}$ to $j^{th}$. In our case these species are the metallofullerenes obtained in the reactions between the metal (M) and C$_{60}$:
\begin{equation*}
  \mathrm{M + C_{60} \rightarrow [M@C_{60}]\; or\; [M-C_{60}]}
\end{equation*}
\begin{equation*}
  \mathrm{M^{+} + C_{60} \rightarrow [M@C_{60}]^{+}\; or\; [M-C_{60}]^{+}}
\end{equation*}
\begin{equation*}
  \mathrm{M + C_{60}^{+} \rightarrow [M@C_{60}]^{+}\; or\; [M-C_{60}]^{+}}
\end{equation*}
We note that the third reaction implying a metal and $\mathrm{C_{60}^{+}}$ can
be discarded because there is no evidence for the presence of C$_{60}^{+}$ in
the circumstellar environments around PNe and RCB stars
\citep{Berne2013,Cordiner2019}. According to the aforementioned reactions, the
metallofullerenes concentration thus depends on the amount of M, M$^{+}$, and
C$_{60}$. From stellar nucleosynthesis models or direct measurements (see below) is possible
also to extract the total metal abundances M$_{T}$ = M + M$^{+}$, while C$_{60}$
can be estimated as the excess reactant due to the larger abundance of carbon C
$>>>$ M$_{T}$. Fullerene-rich PNe display a narrow  $T_{eff}$ range
($\sim$30,000-50,000 K) and chemical abundances similar to their progenitors; i.e.,
slightly metal-poor (Z$\sim$0.004) and low-mass (1.5-2.5 M$_\odot$) asymptotic giant
branch (AGB) stars \citep{Otsuka2014}. We note that our simulated IR spectra do not change significantly for 1.5-2.5 M$_\odot$, due to the fact that the relative metal abundances remain almost constant in this mass range.  Thus, for C$_{60}$-PNe the metal
abundances for a 2 M$_\odot$ AGB star were taken from AGB nucleosynthesis models \citep{Karakas2010,Karakas2018}; note also that not all metals considered here have been directly measured in C$_{60}$-PNe and we thus need the metal abundance values from theoretical predictions. However, the total
metal abundances measurements reported by \cite{Jeffery2011} were used for the
only two fullerene-rich RCBs V854 Cen ($T_{eff}$=6,750 K) and DY Cen
($T_{eff}$=19,500 K) \citep{Garcia2011a}. Accordingly, M and M$^{+}$ are the
limiting factors in the metallofullerenes formation. Using the well-know Saha
equation \citep{Saha1920} we are able to predict the concentration of M and M$^{+}$ by:
\begin{equation}\label{eqn:2}
  \frac{n_{M^{+}}^{2}}{n_{M}}=\frac{G_{M^{+}}g_{e}}{G_{M}}\;\frac{(2\pi m_{e}kT)^{3/2}}{h^{3}}\;\mathrm{exp}\left(-\frac{IP_{1}}{kT_{eff}}\right)
\end{equation}
with $n_{M^{+}}$ and $n_{M}$ being the density of atoms in ionized and neutral states, respectively. The degeneracy of the ionized or neutral state is represented by $G_{M^{+}}$ or $G_{M}$ whilst $g_{e}$ is the electron degeneracy. Finally, $T_{eff}$ is the effective temperature of the PNe/RCB central star, and $IP_{1}$ denotes the first ionization potential of the metal; remained variables are the Boltzmann ($k$) and Planck ($h$), and the electron mass $m_{e}$ constants. Applying also $n_{M_{T}}=n_{M^{+}}+n_{M}$ and $\chi = \dfrac{n_{M^{+}}}{n_{M_{T}}}$, where $\chi$ indicates the fraction of ionized atoms, we obtain:
\begin{equation}\label{eqn:3}
  \frac{\chi^{2}}{1-\chi}=\frac{\dfrac{G_{M^{+}}g_{e}}{G_{M}}\;\dfrac{(2\pi m_{e}kT)^{3/2}}{h^{3}}\;\mathrm{exp}\left(-\dfrac{IP_{1}}{kT_{eff}}\right)}{n_{M_{T}}}
\end{equation}
From equation \ref{eqn:3} we extract $n_{M^{+}}$ and $n_{M}$ to weight the intensities according to:
\begin{equation}\label{eqn:4}
  I_{M^{+}}\propto\sum^{j}_{i}\left(\frac{\partial \mu}{\partial x}\right)^{2}n_{M^{+}}
\end{equation}
\begin{equation}\label{eqn:5}
  I_{M}\propto\sum^{j}_{i}\left(\frac{\partial \mu}{\partial x}\right)^{2}n_{M}
\end{equation}
\begin{equation}\label{eqn:6}
  I_{T} = I_{M} + I_{M^{+}}
\end{equation}
The variable $I_{T}$ is used to construct the total weighted mixture spectra of metallofullerenes and $I_{M}$,$I_{M^{+}}$ the corresponding neutral and charged contributions.

\bibliography{Metallofullerenes_dagh}{}
\bibliographystyle{aasjournal}

%


\end{document}